\documentclass[aps,twocolumn,amsmath,amssymb,showpacs,prl,superscriptaddress,unsortedaddress]{revtex4}
\usepackage{epsf}
\usepackage{graphicx}

\newcommand{\etal}{{\it et al.}}

\begin{document}

\title{Anomalies in the Fermi surface and band dispersion of quasi-one-dimensional CuO chains in
   the high-temperature superconductor YBa$_2$Cu$_4$O$_8$}

\author{Takeshi Kondo}
\affiliation{Ames Laboratory and Department of Physics and Astronomy, Iowa State University, Ames, IA 50011, USA}

\author{R.~Khasanov} 
\affiliation{Laboratory for Muon Spin Spectroscopy, Paul Scherrer Institute, CH-5232 Villigen PSI, Switzerland}

\author{J. Karpinski}
\affiliation{Laboratory for Solid State Physics ETH Z\"urich, CH-8093 Z\"urich, Switzerland}

\author{S. M. Kazakov}
\affiliation{Laboratory for Solid State Physics ETH Z\"urich, CH-8093 Z\"urich, Switzerland}

\author{N. D. Zhigadlo}
\affiliation{Laboratory for Solid State Physics ETH Z\"urich, CH-8093 Z\"urich, Switzerland}

\author{Z. Bukowski}
\affiliation{Laboratory for Solid State Physics ETH Z\"urich, CH-8093 Z\"urich, Switzerland}

\author{M. Shi}
\affiliation{Swiss Light Source, Paul Scherrer Institute, CH-5232 Villigen PSI, Switzerland}

\author{A.~Bendounan}
\affiliation{Laboratory for Neutron Scattering, ETH Z\"urich and Paul Scherrer Institute, CH-5232 Villigen PSI, Switzerland}

\author{Y.~Sassa}
\affiliation{Laboratory for Neutron Scattering, ETH Z\"urich and Paul Scherrer Institute, CH-5232 Villigen PSI, Switzerland}

\author{J. Chang}
\affiliation{Laboratory for Neutron Scattering, ETH Z\"urich and Paul Scherrer Institute, CH-5232 Villigen PSI, Switzerland}

\author{S. Pailh\'es}
\affiliation{Laboratory for Neutron Scattering, ETH Z\"urich and Paul Scherrer Institute, CH-5232 Villigen PSI, Switzerland}

\author{J. Mesot}
\affiliation{Laboratory for Neutron Scattering, ETH Z\"urich and Paul Scherrer Institute, CH-5232 Villigen PSI, Switzerland}

\author{J.~Schmalian}
\affiliation{Ames Laboratory and Department of Physics and Astronomy, Iowa State University, Ames, IA 50011, USA}

\author{H. Keller}
\affiliation{Physik-Institut der Universit\"{a}t Z\"{u}rich,
Winterthurerstrasse 190, CH-8057 Z\"urich, Switzerland}

\author{A. Kaminski}
\affiliation{Ames Laboratory and Department of Physics and Astronomy, Iowa State University, Ames, IA 50011, USA}

\date{\today}
\begin{abstract}
We have investigated the electronic states in quasi one dimensional (1D) CuO chains by microprobe Angle Resolved Photoemission Spectroscopy ($\mu$ARPES). We find that the quasiparticle Fermi surface consists of six disconnected segments, consistent with recent theoretical calculations that predict the formation of narrow, elongated Fermi surface pockets for coupled CuO chains. In addition, we find a strong renormalization effect with a significant kink structure in the band dispersion. The properties of this latter effect [energy scale ($\sim$40 meV), temperature dependence and behavior with Zn-doping] are identical to those of the bosonic mode observed in CuO$_2$ planes of high temperature superconductors, indicating they have a common origin. 
\end{abstract}


\maketitle 
In one dimensional (1D) systems, electrons are confined to motion along the atomic chains making them strongly correlated \cite{VOIT}.  The electronic properties of 1D CuO chains are especially interesting because they form the building blocks for CuO$_2$ planes, that displays high temperature superconductivity. The pairing mechanism in these systems is not well understood. One reason for this is the difficulty in modeling of CuO$_2$ planes. A possible solution is to start with very weakly coupled CuO chains, which are well understood analytically, and then increase the coupling to a value typical for CuO$_2$ planes. Recently, a calculation \cite{Giamarchi2} for coupled CuO chains was performed as a function of coupling strength using a variation of the Dynamical Mean Field Theory (ch-DMFT) \cite{Arrigoni,Georges,Giamarchi1}. It revealed the breakdown of a typical quasi-1D Fermi surface into narrow, elongated pockets that resemble the disconnected Fermi arcs observed by ARPES in the pseudogap state of the cuprates \cite{Norman_nature}. This issue is very important as a number of recent experiments find quantum oscillations in YBa$_2$Cu$_3$O$_{7-\delta}$ (Y123) and YBa$_2$Cu$_4$O$_8$ (Y124), which suggests the existence of very small Fermi pockets \cite{Taillefer,Cooper,Hussey,Hardy}.  A detailed understanding of the physics of 1D CuO chains is also important because they provide an idealized model for stripes that are seen in some cuprates and which are thought by some to play a role in the mechanism of high temperature superconductivity \cite{Tranquada1,Tranquada2,Kivelson1,Kivelson2,Kivelson3}. 
Surprisingly, the signature of magnetic fluctuations in stripes \cite{Tranquada3} very closely resembles  that associated with the magnetic resonant mode seen in cuprates by inelastic neutron measurements\cite{Hayden}. Therefore the electronic properties of chains may also provide important information about relation between magnetic resonant mode, high temperature superconductivity and stripes.

In this letter, we use angle resolved photoemission spectroscopy and study the electronic properties of CuO chains in the high temperature superconductor, YBa$_2$Cu$_4$O$_8$ (Y124). 
We observe a quasiparticle Fermi surface that consists of disconnected segments. 
This observation is in excellent agreement with, and directly confirms, the results of the ch-DMFT calculation for coupled chains\cite{Giamarchi2}.  The result raises the possibility that the quantum oscillations observed in Y123 and Y124 \cite{Taillefer,Cooper,Hussey,Hardy} occur in pockets due to CuO chains.
In addition, our data reveals the presence of a strong renormalization effect at an energy of $\sim$40 meV and a small energy gap ($\sim$5meV) in the superconducting state. 
This effect is only observed along a small momentum region, where the Fermi surface of  chains  and  planes  crosses, suggesting that it is caused by coherent hopping of electrons between the chains and planes. 
The renormalization effect disappears above $T_c$ and is almost completely suppressed in samples doped with only 1$\%$ of Zn. These properties and the energy scale ($\sim$40 meV) are almost identical to those of the bosonic mode observed in superconducting CuO$_2$ planes \cite{Kaminski,Sato,Terashima,Mook,Fong,Sidis}, which suggests they have a common origin. 
Our data demonstrates that chains play a crucial role in establishing a 3D metallic state and superconductivity  in YBCO at low temperatures \cite{Hussey_resistivity1,Hussey_resistivity2}.

\begin{figure}
\includegraphics[width=3.in]{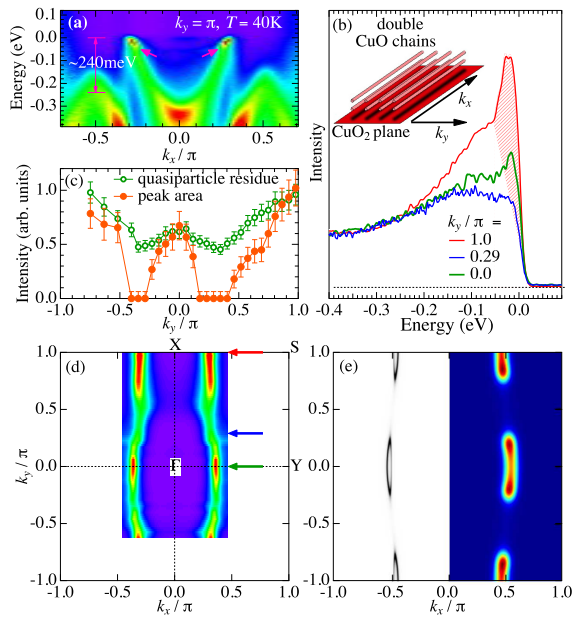}
\caption{(Color online)
(a) Band dispersion map along the chains. ($h\nu$ = 22eV) (b) EDCs at three  $k_F$s indicated by arrows in (d). (c) Quasiparticle residue (spectral intensity integrated within $\pm 50$meV of $E_F$), and area of the quasiparticle peak (illustrated in (b)), along the Fermi surface.   (d) Measured Fermi surface determined by integrating the EDCs within $\pm 50$meV of $E_F$ at $T$ = 20K with $h\nu$ = 33eV. (e) Calculated Fermi surface\cite{Giamarchi2}: zero-energy spectral function $A(k,0)$ at $T=0$K (left image) and expected ARPES intensity $I(k,0)$  (right image).}  
\label{fig1}
\end{figure}

Twin-free single crystals of underdoped pristine YBa$_{2}$Cu$_{4}$O$_{8}$ ($T_c=80$K) and 1$\%$ Zn-doped YBa$_{2}$Cu$_{4}$O$_{8}$ ($T_c=60$K), were grown by the self-flux method under high oxygen pressure\cite{Karpinski}.   ARPES measurements were performed at the PGM beam line at the Synchrotron Radiation Center (SRC), USA, and at the SIS beamline of Swiss Light Source (SLS), Switzerland. Both measurements used a SES2002 hemispherical analyzer. The energy and angular resolutions were $10-30$meV and $\sim
$0.15$^{\circ}$, respectively.  

\begin{figure}
\includegraphics[width=3.in]{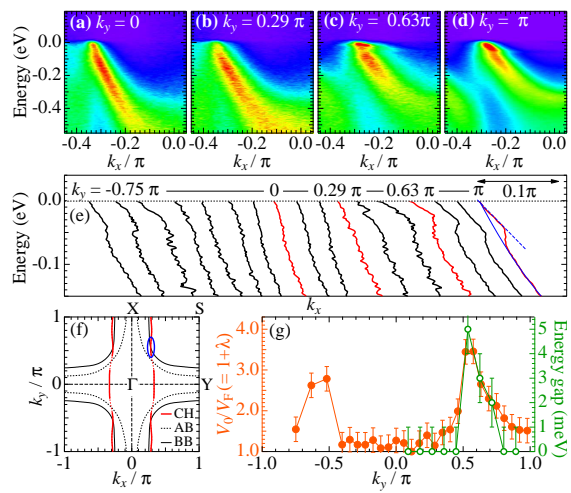}
\caption{(Color online) 
(a-d) Band dispersion maps along chains. ($T$=20K, $h\nu$=33eV) (e) Band dispersion plots determined from MDC peak positions. (f) Fermi surface of chains (CH) and plains (AB: antibonding band, BB: bonding band). The superconducting gap region is marked by blue circle. (g) Ratio between quasiparticle velocities without and with the mode coupling ($V_0/V_F$), and superconducting gap along Fermi surface.  The gap size was estimated from high resolution data at $h\nu$=22eV.} 
\label{fig2}
\end{figure}

Figure 1(a) shows a typical band dispersion map from the CuO chains. A schematic diagram of the chain domains at the sample surface is shown in the inset of Fig. 1(b). As we reported in a previous paper\cite{Kondo}, two bands originating from CuO chains are clearly seen: the higher energy band is conducting while the low energy band is insulating. The conduction band most likely originates from one of the double chains furthest from the surface and its carrier concentration and electronic properties are closer to that of the bulk, where the chains are known to be metallic \cite{Hussey_resistivity1,Lee}.  In contrast, the gapped insulating band is most likely due to the chain layer that is exposed to the surface. It is almost isolated from the rest of the system and has a very different carrier concentration. In the rest of this letter, we present intriguing properties of the conduction band. 

In Fig. 1(d), we plot our measured Fermi surface from the conducting CuO chains. A fascinating feature is the strong variation of the spectral intensity along the 1D-like Fermi surface. To study this further, we examine in Fig. 1(b) the typical energy distribution curves (EDCs) at three $k_F$ points along momentum cuts indicated in the Fig. 1 (d) with arrows.  We find that the quasiparticle peak is absent in some regions (blue curve in Fig. 1 (b)).  We estimate the spectral weight of the quasiparticle peak (shaded area in Fig. 1 (b)) and plot it in Fig. 1 (c) along with the intensity at $E_F$, which represents the quasiparticle residue. The absence of a quasiparticle peak produces six disconnected segments of a quasiparticle Fermi surface in momentum space (islands of strong intensity in Fig. 1 (d)). This result is in excellent agreement with a recent ch-DMFT calculation\cite{Giamarchi2}, which predicts thin elongated Fermi pockets in coupled CuO chains. The left image of Fig.1 (e) shows the result of this calculation. The authors of ref.\cite{Giamarchi2} also suggest in the right half of Fig. 1 (e) what these pockets would look like in ARPES, with its finite energy and momentum resolution. The pockets are so narrow that the ARPES intensity at the Fermi level is expected to appear as six arc-like segments - as we indeed see in our data. The location of the calculated islands ($k_x$ value) is different from our data: the calculation assumes half-filling and the experimental situation is close to quarter-filling. Here it is worth mentioning that the calculation assumes a spinless situation and the model can be viewed as a caricature of a spin-1/2 model with quarter-filling and a very strong local repulsion\cite{Giamarchi2}. 

\begin{figure}
\includegraphics[width=3.in]{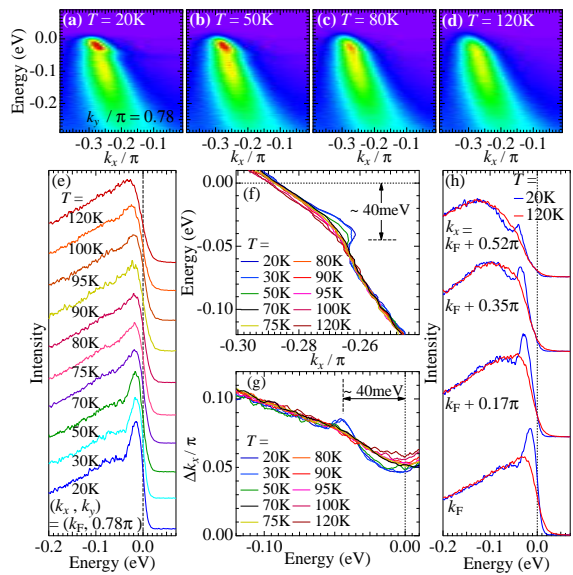}
\caption{(Color online) 
(a-d) Band dispersion maps at $k_y/\pi=0.78$ measured at various temperatures from deep below to above $T_c$. ($h\nu$ = 33eV) (e) EDCs at $k_F$, (f) Band dispersions determined from MDC peak positions, and (g) MDC widths at various temperatures. (h) EDCs close to $k_F$ below and above $T_c$. }
\label{fig3}
\end{figure}

 Our results for CuO chains in Y124 are naturally more complex than those predicted by the ch-DMFT calculations. In particular we find a significant band renormalization effect near the Fermi level that appears as a ``kink" in dispersion (see arrows in Fig.1(a)) most likely due to a coupling between the CuO chain electrons and a collective mode associated with chains. 
We confirmed that the energy dispersion with the kink has only a single peak in MDC at $E_F$. The energy of the``dip'' in EDC associated with the kink structure has constant binding energy regardless of the momentum. 
These features are all consistent with the bosonic mode interpretation for the kink and exclude the possibility that two overlapping bands are the culprit.
Furthermore, the photon energy studies of this feature (not shown) reveal that the intensities of the low and high binding energy portion of this feature have the same photon energy dependence.  If this feature would originate from two different bands, their intensities would have different photon energy dependence due to different matrix element effects.

\begin{figure}
\includegraphics[width=3in]{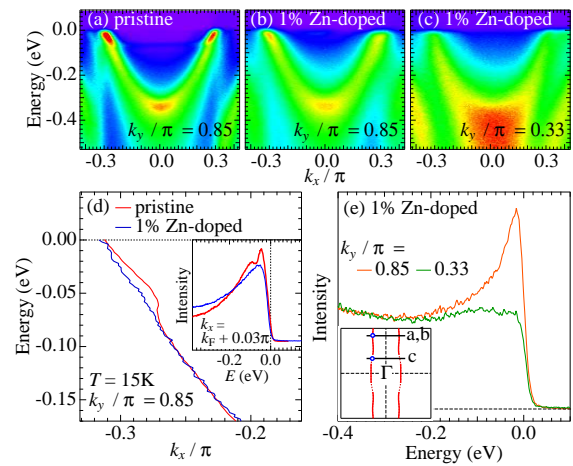}
\caption{(Color online) 
Band dispersion maps for (a) pristine and (b-c) 1 $\%$ Zn-doped sample. The corresponding momentum cuts are indicated in inset of (e)  (d) Band dispersion plots determined from MDC peak positions for (a) and (b). Inset shows EDCs slightly off $k_F$.
(e) EDCs at $k_F$s for (b) and (c). }
\label{fig4}
\end{figure}

Now we need to clarify whether or not there is a relationship between the quasiparticle peaks and the mode coupling, which may enhance the peak intensity. Figure 2(a)-(d) shows the band dispersion along the chains measured at several values of $k_y$ below $T_c$. The strength of the kink structure in the dispersion changes dramatically with $k_y$ momentum. The momentum variation is seen more clearly in Fig. 2 (e), where we plot the band dispersion close to $E_F$.  
We estimated the strength of the renormalization by taking the ratio of the quasiparticle velocity at the Fermi level without and with the mode coupling $V_0/V_F$, and plot it in Fig. 2 (g). 
$V_F$ is simply the slope of the dispersion data. $V_0$ was estimated as the slope of the parabolic band extrapolated from high energy.
Moving from the zone boundary to the center, we find that the ratio ($V_0/V_F$) increases sharply in a narrow momentum range and then drops to almost one (orange circles, Fig. 2 (g)).  This momentum variation is very different from that of the quasiparticle peaks (Fig. 1(c)), indicating that the quasiparticle peaks have no direct relation with the mode coupling. 
We note that the renormalization effect becomes significant in a region of the Brillouin zone where the chain and plane Fermi surfaces cross each other (see Fig. 2(f)). In addition, we find the  superconducting gap which varies strongly with momentum (green open circles, Fig. 2 (g)) following the strength of renormalization. The direct relationship between the strength of the renormalization and the value of energy gap is similar to that observed in superconducting CuO$_2$ planes, where these two parameters are maximal at the antinode\cite{Kaminski,Sato,Terashima}.  
It seems that electrons in the chains and planes significantly interact with each other via the bosonic mode in a small momentum region\cite{Lee,Atkinson}. The  superconducting gap ($\sim$5meV) observed in the chains is significantly smaller than that in planes\cite{Nakayama}. This might be because the superconducting gap in the CuO chains is due to a proximity effect originating from the planes\cite{Kohen}. 
It was reported that $c$-axis conductivity remains metallic at low temperatures in Y124 in the underdoped regime.\cite{Hussey_resistivity1,Hussey_resistivity2}  
This is in contrast to other underdopded cuprates, which show insulating behavior along the $c$-axis.
Furthermore, YBCO compounds are
characterized by relatively strong three-dimensional
superconductivity.
Our data strongly suggests that nontrivial electron hopping between the chains and planes plays a significant role in enhancing 3D properties of this most commonly studied high temperature superconductor.

We investigate the properties of the renormalization effect in more detail.  Figure 3 (a)-(d) shows the band dispersion at $k_y=0.78\pi$ measured at various temperatures below and above $T_c$.  The EDCs at $k_F$, band dispersion and momentum distribution curves (MDCs) widths (related to the scattering rate of the electrons) are plotted as a function of temperature in Fig. 3 (e), (f) and (g), respectively. The renormalization effect is clearly seen at $\sim$40 meV below $T_c$ and is suppressed above $T_c$, so the effect is tied to the superconductivity. In Fig. 3 (h), it is demonstrated that the characteristic peak-dip-hump structure in EDCs close to $k_F$ disappears above $T_c$. 
The temperature dependence and energy scale of the renormalization effect are similar to those of the bosonic mode observed in superconducting CuO$_2$ planes.  
 
To elucidate the origin of the observed collective mode we measured the band dispersion in 1$\%$ Zn-doped Y124 samples, and compared the one for pristine samples. The addition of Zn drastically changes the magnetic environment with no significant effect on the lattice vibrations because of the similar masses of Zn and Cu. ARPES data on pristine and Zn-doped samples measured under exactly the same conditions are shown in Fig.4 (a) and (b), respectively. The corresponding band dispersion and EDCs slightly off the $k_F$ are compared in Fig. 4 (d). 
Although the entire band shape (bottom energy and $k_F$ value) are almost identical in the both samples,  
the kink structure and the peak-dip-hump spectral shape observed in the pristine samples are effectively absent in Zn-doped samples.
This strongly suggests that the  renormalization effect in the CuO chains has magnetic origin and it is due to coupling between the electrons and spin fluctuations. 
A similar  Zn-doping effect is observed in superconducting CuO$_2$ planes\cite{Terashima} strongly supporting the view that the bosonic mode has a common origin in CuO chains and CuO$_2$ planes. 

Figure 4(b) and (c) shows the band dispersion measured along two momentum cuts, where spectra with and without quasiparticle peaks are observed in pristine samples, respectively. The corresponding EDCs at $k_F$s are plotted in Fig. 4 (e). Clearly the kink is absent from the Zn-doped samples (Fig. 4(d)), but the quasiparticle peaks remain in the same regions of the Brillouin zone as the pristine samples. This result supports our idea that the quasiparticle peaks are not due to a collective mode, but are due to chain coupling as predicted by the ch-DMFT calculations.

In conclusion, we observed three intriguing features in CuO chain band of YBa$_2$Cu$_4$O$_8$. 
One is a quasiparticle Fermi surface that consists of six disconnected segments.
This is in excellent agreement with the results of ch-DMFT calculations for coupled 1D chains. 
Second  is a strong renormalization effect. 
The properties of this effect are very similar to those of the bosonic mode observed in the CuO$_2$ planes. As a third significant result, we observed a small superconducting gap in CuO chains that is most likely due to proximity effect originating from the planes.

We thank C. Berthod and T. Giamarchi for discussions and Fig. 1(e).
This work was supported by Director Office for Basic Energy Sciences, US DOE and Swiss NCCR MaNEP. The Ames Laboratory is operated for the US DOE by Iowa State University under Contract No. W-7405-ENG-82. The Synchrotron Radiation Center is supported by NSF DMR 9212658. R. K. gratefully acknowledges support of Swiss National Science Foundation and K. Alex M\"uller Foundation.


\begin{thebibliography}{99}
\bibitem{VOIT}
J. Voit, El. Prop. Nov. Mat. {\bf 544}, 309 (2000).

\bibitem{Giamarchi2}
C. Berthod,  T. Giamarchi, S. Biermann and A. Georges, Phys. Rev. Lett.  {\bf 97}, 136401 (2006).

\bibitem{Arrigoni}
E. Arrigoni, Phys. Rev. Lett. {\bf 83}, 128 (1999).

\bibitem{Georges}
A. Georges, T. Giamarchi, and N. Sandler, Phys. Rev. B {\bf 61}, 16393 (2000).

\bibitem{Giamarchi1}
S. Biermann, A. Georges, A. Lichtenstein, and T. Giamarchi, Phys. Rev. Lett.  {\bf 87}, 276405 (2001).

\bibitem{Norman_nature}
M. R. Norman $et\ al$., 
Nature {\bf 392}, 157 (1998).

\bibitem{Taillefer}
Nicolas Doiron-Leyraud {\etal}, Nature {\bf 447}, 565 (2007).

\bibitem{Cooper}
E. A. Yelland {\etal}, Phys. Rev. Lett.  {\bf 100}, 047003 (2008).

\bibitem{Hussey}
A. F. Bangura {\etal}, Phys. Rev. Lett.  {\bf 100}, 047004 (2008).

\bibitem{Hardy}
Cyril Jaudet {\etal}, Phys. Rev. Lett.  {\bf 100}, 187005 (2008).

\bibitem{Tranquada1}
J. M. Tranquada, B. J. Sternlieb, J. D. Axe, Y. Nakamura, and S. Uchida, Nature {\bf 375}, 561 (1995).

\bibitem{Tranquada2}
J. M. Tranquada {\etal},  Phys. Rev. Lett. {\bf 78}, 338 (1997).

\bibitem{Kivelson1}
S. A. Kivelson, E. Fradkin and V. J. Emery, Nature {\bf 393}, 550 (1998).

\bibitem{Kivelson2}
E. W. Carlson, D. Orgad, S. A. Kivelson, and V. J. Emery, Phys. Rev. B. {\bf 62}, 3422 (2000).

\bibitem{Kivelson3}
E. Berg, T. H. Geballe and S. A. Kivelson, Phys. Rev. B. {\bf 76}, 214505 (2007).

\bibitem{Tranquada3}
J. M. Tranquada \etal, Nature {\bf 429}, 534-538 (2004) 

\bibitem{Hayden}
S. M. Hayden \etal, Nature {\bf 429}, 531-534 (2004) 

\bibitem{Kaminski}
A. Kaminski  {\etal}, Phys. Rev. Lett.  {\bf 86}, 1070 (2001).

\bibitem{Sato}
T. Sato {\etal}, Phys. Rev. Lett.  {\bf 91}, 157003 (2003).

\bibitem{Terashima}
K. Terashima {\etal}, Nature physics {\bf 2}, 27 (2006).

\bibitem{Mook}
H. A. Mook, M. Yethiraj, G. Aeppli, T. E. Mason and T. Armstrong, Phys. Rev. Lett. {\bf 70}, 3490 (1993).

\bibitem{Fong}
H. F. Fong {\etal}, Nature {\bf 2}, 27 (1999).

\bibitem{Sidis}
Y. Sidis {\etal}, Phys. Rev. Lett. 84, 5900 (2000).

\bibitem{Hussey_resistivity1}
N. E. Hussey, K. Nozawa, H. Takagi, S. Adachi and K. Tanabe, Phys. Rev. B {\bf 56}, R11423 (1997).

\bibitem{Hussey_resistivity2}
N. E. Hussey, M. Kibune, H. Nakagawa, N. Miura, Y. Iye, H. Takagi, S. Adachi and K. Tanabe, Phys. Rev. Lett. {\bf 80}, 2909 (1998).

\bibitem{Karpinski}
J. Karpinski $et\ al$.,  
Nature {\bf 336}, 660 (1988); Supercond. Sci. Technol. {\bf 12}, R153 (1999).

\bibitem{Kondo}
T. Kondo $et\ al$., 
Phys. Rev. Lett. {\bf 98}, 157002 (2007).

\bibitem{Lee}
Y.-S Lee, K. Segawa, Y. Ando, and D. N. Basov, Phys. Rev. Lett. {\bf 94}, 137004 (2005).


\bibitem{Atkinson}
W. A. Atkinson, Phys. Rev. B. {\bf 59}, 3377 (1999).

\bibitem{Nakayama}
K. Nakayama {\etal}, Phys. Rev. B {\bf 79}, 140503(R) (2009).

\bibitem{Kohen}
A. Kohen, G. Leibovitch, and G. Deutscher, Phys. Rev. Lett. {\bf 90}, 207005 (2003).

\end{thebibliography}
\end{document}